\documentclass{kapproc} 

\usepackage{procps}

\usepackage[dvips]{graphicx}

\upperandlowercase

 \let\footnote\savefootnote

\normallatexbib 

\begin{document}

\articletitle
{HI Clouds Beyond the Galactic Disk}

\chaptitlerunninghead{Halo Clouds} 

\author{Felix J. Lockman}

\affil{National Radio Astronomy Observatory, Green Bank,
WV USA}

\begin{abstract}
Recent observations in the 21cm line with the Green Bank Telescope 
have changed our view of the neutral interstellar 
medium (ISM) in several ways.  The new data show that in 
 the inner parts of the Milky Way the disk-halo interface is 
 composed of many discrete HI clouds. The clouds
  lie in a layer more than one kpc thick and  follow Galactic rotation. 
Their origin and evolution is unknown.  
In the outer Galaxy, the new data show that 
the high-velocity cloud Complex H is likely a satellite on a 
retrograde orbit interacting with some 
extended component of the Milky Way's ISM.  
  These observations place new constraints on models of 
the ISM and are directly related to the work of 
  Don Cox and Ron Reynolds.
\end{abstract}

\section{Introduction}

This paper is  about two topics  close to the 
 interests of  Don Cox and Ron Reynolds: 
 the structure of the ISM at the disk-halo interface, and 
a high-velocity HI cloud which appears to be interacting 
with the gaseous Galactic halo.  The discoveries discussed here 
were made with 
the new 100 meter Green Bank Telescope (the GBT), whose 
 sensitivity, dynamic range, and angular resolution 
 make it a fabulous tool for Galactic 21cm HI observations.  
We are just now beginning to feel its impact in interstellar and 
Galactic studies.

\begin{figure}[ht]
\centerline{\includegraphics[height=1.25\hsize]{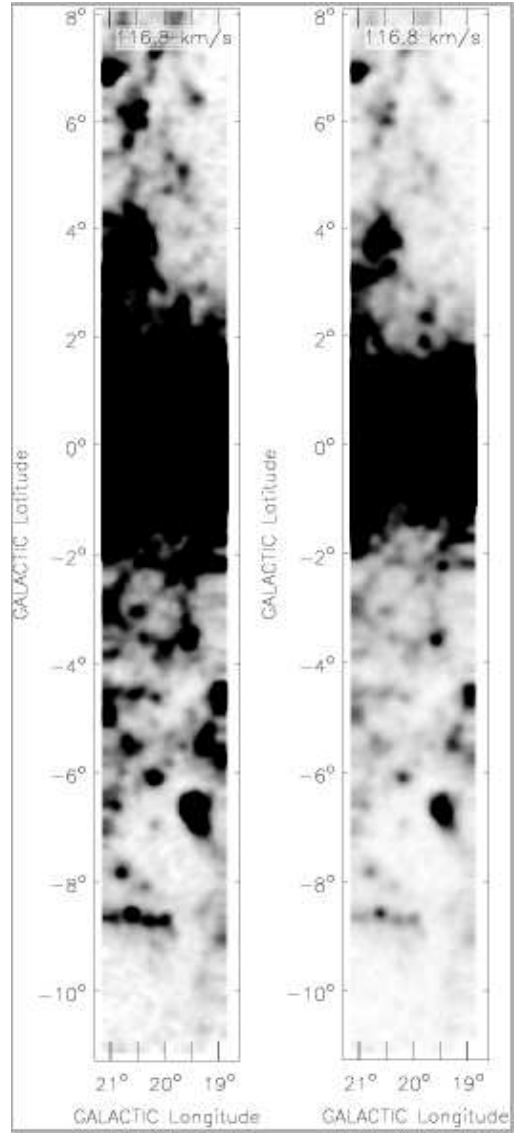}}
\caption{GBT observations of  HI near longitude $20^{\circ}$ 
 at $V_{LSR} = 116.8$ km s$^{-1}$. 
The two panels show identical data  plotted on  different 
intensity scales: the left panel to emphasize fainter emission, the 
right panel to emphasize the brighter.  
 The angular resolution of the observations is $9'$ and 
observations were made on a $3'$ grid with an integration time 
of 5 seconds per point.  At the velocity shown here one degree in 
either coordinate corresponds to a linear scale of about 140 pc. 
Some of the clouds are more than one kpc from the Galactic 
midplane, and cloud-like structures persist quite far down 
toward the disk.
  }
\end{figure}

\section{HI That Came Up From the Disk}

It has been known for some time that there  are significant amounts
of HI far from the Galactic disk that 
 can be observed in 21cm emission and 
optical/UV absorption lines  (e.g., Lockman 1984; 
Lockman, Hobbs \& Shull 1986;  Savage 1995). 
Until now, however, its structure has been unknown.  
Figure 1 shows GBT observations of the  HI in the Galactic plane and 
lower halo  around longitude $20^{\circ}$. In these new data 
the transition zone between the neutral disk 
and the halo is seen to be populated with neutral clouds.  
A typical halo cloud in a sample studied near $\ell = 29^{\circ}$ 
(Lockman 2002) has a size of a few tens of pc, an HI mass of 
a few tens of solar masses, and  $N_{HI} = 2 \times 10^{19}$ 
cm$^{-2}$, though clouds have a large range in all of their properties.    
The HI lines from the clouds have a typical width 
of 12 km s$^{-1}$, but some lines are so narrow that  components 
within the clouds must have $T < 1000$ K.  It is common to find 
halo clouds with two line components  --- one broad 
and one narrow --- at the same velocity, 
implying that there is HI at two distinct 
temperatures.   This state is possible for diffuse HI at some pressures 
(Field, Goldsmith \& Habing 1969). 

These clouds have kinematics which are dominated by Galactic 
rotation even when they are more than 1 kpc from the plane.  Their 
connection to events in the disk thus seems secure.  Yet to what are 
they connected?  Are they formed from neutral gas thrust upward 
 by supernovae (e.g. Heiles 1984; 
Norman \& Ikeuchi 1989; de Avillez \& Berry 2002)?
  Or are they the return products of a Galactic fountain: cool clouds 
condensing from the  very 
 hot gas in the halo (Shapiro \& Field 1976)?  In either case 
the clouds consist of material that began in the disk.

The HI clouds are denser than their surroundings by 
orders of magnitude so  they must be falling toward the plane 
like rocks: the  free-fall time is 
$\sim 50$ Myr.  Yet they are fairly diffuse objects, and have too 
little mass in HI to be bound gravitationally. 
Unless confined,   these clouds will dissipate on a time scale 
$Diam/ \Delta v \sim 2 $ Myr.  The halo clouds may be in  pressure 
equilibrium with the Reynolds Layer of H$^+$ or with 
the halo of very hot gas.  
At 1 kpc from the plane 
the average densities $\langle n(HI) \rangle  \approx 
\langle n(H^{+}) \rangle$ within a factor of two (Dickey \& Lockman 1990; 
Reynolds 1997) but both species probably have a small filling factor.
Magnetic fields may have a role in the maintenance of the 
halo clouds, as they may for high-velocity 
clouds (Konz, Br\"uns \& Birk 2002).

Study of the  HI clouds  is just beginning, and only a few 
 have been observed at high angular resolution. 
In  the existing data 
 the halo clouds do not look as if they  are simply the denser 
peaks in a continuous  medium but 
appear to be isolated objects, and I suspect that they are stable 
for periods larger than their sound-crossing time.  They seem to be 
genuine interstellar ``clouds''.

\subsection{Is the long search for interstellar clouds finally over?}

For 50 years the diffuse ISM has been described as 
containing `clouds':  discrete objects with distinct 
boundaries in position and velocity (e.g., Munch 1952).  
Yet such clouds have never been observed.  
The only structures seen in HI emission which fit that 
description have  peculiar 
velocities --- the high- and intermediate-velocity clouds --- and 
are not associated with the Galactic disk.  HI emission studies 
sometimes find blended filaments and portions of sheets 
(Kulkarni \& Heiles 1988), but no  clouds. 

Now, in the GBT data, there are hundreds  of diffuse clouds, 
 though most  are far from the Galactic plane.  
Has the GBT finally revealed the fundamental structure of the ISM  
which has been 
 hidden from  previous generations of instruments, or 
is this a population strictly confined to the halo?   
The data suggest that the situation is not so simple.  The 
HI in our part of the Galaxy --- the ISM we look through when doing 
most extragalactic astronomy ---  does not appear extremely lumpy.
Clouds of the type seen in Fig.~1 would be quite obvious 
even in older data if  some were located  at high latitude near the Sun. 
It is  possible that 
we are being mislead,  but  local HI just does not seem to 
decompose into clouds as  halo HI in the inner Galaxy (or some fraction of it) 
 does.  And yet, as Fig.~1 clearly shows, in the inner Galaxy clouds 
are observed to fairly low latitudes where they appear to blend 
together into an indistinguishable mass.  Lockman \& Stil (2004) 
present an example of a discrete cloud in the Galactic plane which 
is observable only because it has a high random velocity and 
thus lies unconfused in the wing of the HI spectrum.  
It resembles a McKee \& Ostriker (1977) cold cloud core with 
a peculiar velocity $\sim 50$ km s$^{-1}$. 
It may  be an example of the ``fast'' HI clouds postulated to 
explain the  wings of HI profiles in the Galactic plane 
(Radhakrishnan \& Srinivasan 1980; Kulkarni \& Fich 1985).  
But it is a diffuse cloud nonetheless, that looks like a ``halo'' cloud, 
 and lies only 14 pc from $b=0^{\circ}$. I think that 
the  ISM may have a different structure locally than in the  inner 
Galaxy, where HI clouds with a high cloud-cloud velocity may pervade 
the disk and the halo.

\begin{figure}[ht]
\centerline{\includegraphics[height=0.75\textwidth]{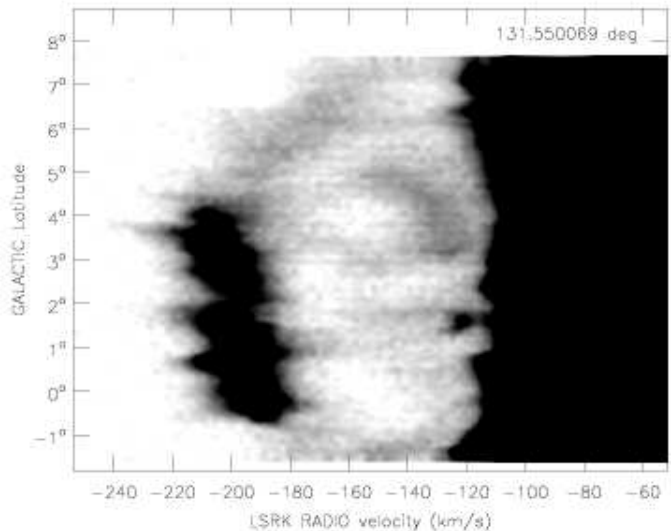}}
\caption{
The HI emission from high-velocity cloud Complex H as measured by 
the GBT.  This  velocity-latitude cut  at $\ell = 131.55^{\circ}$
shows the cloud with a  velocity near $-200$ km s$^{-1}$, 
its velocity gradient with 
latitude, and diffuse emission connecting the cloud 
kinematically to the  hydrogen in 
the Galactic disk HI at the right of the Figure.
}
\end{figure}

\section{HI Coming Down Onto the Disk}

Now we move from gas in normal circular rotation to gas with 
decidedly unusual kinematics, though I will argue that the 
kinematics are not pathological. 
High-velocity HI clouds cover more than one-third of the sky 
(Wakker \& van Woerden 1997).  By 
definition, their velocities cannot be attributed to 
normal Galactic rotation, 
but this does not mean that their velocities  carry no information 
on their origin and fate.  Complex H, in particular, 
 has a large, organized core, lies in the Galactic 
plane (so its peculiar motion 
cannot be attributed entirely to infall),  and 
covers such a large angle that projection effects can be exploited.

Figure 2 shows a velocity-latitude cut through the 
brighter parts of Complex H at longitude $131.5^{\circ}$.  
The figure is overexposed to bring out two key features: (1) the core 
of the complex has a  slope $dV_{LSR}/db = -3 $ km s$^{-1}$ 
deg$^{-1}$ where it crosses the Galactic plane, and (2) 
there is faint HI emission at velocities between the Complex and 
the normal disk, connecting the Complex, kinematically, to the 
Galaxy.  

The velocity gradient is a general feature of the Complex, and is 
important,  for 
although the vertical motion of any object projects 
to zero velocity LSR where it crosses the Galactic plane, {\it near} 
 the Galactic plane $dV_{LSR}/db = V_z$.  Thus for Complex H 
 the change in $V_{LSR}$ with $b$ is most simply understood 
as the projection of the vertical velocity component of the Complex: 
a gradient of 
 --3 km s$^{-1}$ per degree implies  $V_z = -170$ km s$^{-1}$.  
 Complex H is therefore  an HI cloud moving to 
negative Galactic latitude at a substantial velocity.  But what of 
its azimuthal and radial motions?  Several lines of evidence now 
suggest that the circular rotational velocity of the Milky Way is 
constant at about $V_c = V_0 = 220$ km s$^{-1}$
  out to hundreds of kpc from the 
Galactic center (e.g., Zaritzsky 1999; Bellazzini 2003).  If we assume 
 that  Complex H is on a nearly circular orbit 
with a total velocity of 220 km s$^{-1}$, then both its ``anomalous'' 
 velocity and the velocity gradient follow naturally 
if the orbit is inclined, and {\it retrograde}.

In this model,  Complex H is $33\pm9$ kpc from the Galactic center, 
has an orbital inclination  $\approx 45^{\circ}$, and an overall retrograde 
motion with a total $V = 220$ km s$^{-1}$ (Lockman 2003).  
The model reproduces the 
kinematics of the Complex and the relative location of  HI at 
different velocities (Fig.~2; see also Fig.~1 of Lockman 2003).  
At 33 kpc from the Galactic center the cloud 
 should  be interacting with the extended Galactic disk or halo, 
explaining why  the core of the Complex is shrouded with broad-line 
gas at a velocity appropriate for material stripped  by 
its passage through the Milky Way.  Complex H covers such a large 
area on the sky that further orbital constraints may be found from the 
gradient of its $V_{LSR}$  with longitude: the assumption that 
its orbit is nearly circular  might be tested.  

Complex H  appears to be a satellite of the Milky Way with 
$M_{HI} \geq 6 \times 10^6 M_{\circ}$  
and a size  $ > 10 \times 15$ kpc,  
moving on an inclined, retrograde orbit,  
 which is  now passing through the extended Galactic 
disk (or halo) and being fragmented.  It is one of several 
  examples of high-velocity HI clouds which 
are interacting with a gaseous  component of the Milky Way that must extend 
at least  50 kpc from the Galactic center (Konz et al. 2002; 
Putman et al. 2003).  We are beginning to 
take advantage of the fact that high-velocity HI clouds can be probes of 
 conditions far out in the halo, messy probes, but probes nonetheless.  

\section{Questions for Don \& Ron (and several others)}

This meeting is inspired by the work of Don Cox and Ron Reynolds. 
For years Don has taught us to think clearly about 
interstellar physical processes.  
In a field often  sodden with detail, his work stands out for its 
clarity and  focus on the physical facts, 
and his reviews are full of wisdom (e.g., Cox 1990, 1995, 2000).  
 Ron, the consummate observer, has discovered an entirely 
new component of the ISM, though it is still ignored by many.  
He has persisted in making  sensitive measurements and pointing out 
their implications to an often incredulous community 
 (e.g., Reynolds 1989, 1990; Reynolds et al. 1999). 
I hope that the results presented here will delight and confound Don \& 
 Ron in the same way that their work has delighted and confounded me. 
To this end, and in the spirit of this Tertulia, 
instead of a summary I will present some questions 
raised by the GBT data: 

 1) What produces HI clouds so far from the Galactic disk?

 2) Halo HI clouds are 
 exposed to the extragalactic UV radiation field from above and 
 UV leakage from the Galactic disk below.   
Is there a connection between the halo clouds and the Reynolds layer?

3) What sort of medium is Complex H  encountering 
as it moves along its orbit more than 30 kpc from the Sun? 
Like the halo clouds, it lacks enough HI to be self-gravitating, so 
what holds it together?

4) What is the role of magnetic fields  in the halo 
clouds and  in Complex H? If the 
magnetic field is of fundamental importance (as Don reminds us 
 whenever he gets the chance) 
why can so many of us live comfortable, productive lives while 
ignoring it almost entirely?

5) If ``fast'' HI clouds are common at $b\approx0^{\circ}$ 
in the inner Galaxy, 
is the   ISM there  fundamentally different from the ISM near the Sun? 

6) If  the halo clouds are stable for 
many Myr, how exactly are these turbulent objects confined, even 
if there is an external medium?  

7) Does the  ISM know about dark matter, and 
if so, what does it know?  

I suggest that we petition our gracious hosts at the 
Instituto de Astrofísica de Andalucía 
to reconvene this group  again in Granada in, say, in two years, 
to see what has become of these questions, 
and  once again participate, with Don and Ron, 
 in  fruitful discussions which 
may lead  to their resolution.

\begin{acknowledgments}
I thank Yurii Pidopryhora for comments on the manuscript.
The National Radio Astronomy Observatory is operated by Associated 
Universities, Inc., under a cooperative agreement with the 
National Science Foundation.
\end{acknowledgments}

\section{Discussion}

\noindent
{\it C. Konz:} Do you have any information about the metallicity of 
Complex H which would support the idea that this complex is not 
part of the Galaxy?  \\ 
 
\noindent
{\it Lockman:} There is no information about abundances in Complex H.  
To date it has not been detected in any absorption lines at all. \\

\noindent
{\it R. Benjamin:} You said you have an estimate for the mass loss of 
Complex H.  How long does it have to live assuming it sheds mass at 
the same rate? \\

\noindent
{\it Lockman:} Gas that is decelerated from Complex H by 50 km s$^{-1}$ 
or more will blend in velocity 
with the much brighter Galactic disk emission 
and be almost impossible to detect, 
so my estimates are very uncertain.  That being said, 
 it appears that the mass in the HI tail of the Complex is equal to 
that in the compact core.  I suspect, though,  that its mass loss is 
episodic rather than continuous. \\

\noindent
{\it C. Heiles:} You derive properties of halo clouds 
by assuming spherical 
geometry.  What if the objects are 
 extended cylinders along the line of sight? 
With such cylinders the only locations where you could 
see them are those where the cylinders happen to be lined up.  
Well-known examples of such effects are the optical filaments in the 
Cygnus loop, which are sheets seen edge-on.  In your case, you 
see ``clouds'' connected by ``filaments'' -- just what you'd expect 
if they were filaments twisting in the wind. \\

\noindent
{\it Lockman:} First, the clouds {\it look} quasi-spherical, even those
which we have resolved with the VLA.  Many of them  have well-defined 
edges in 
both position and velocity, though not all.  The other point is that 
the clouds often have  narrow lines 
indicating cool gas.  The HI structures in the halo which look like 
filaments always have broad lines.  Thus I believe that 
most of the halo clouds 
are not simply products of some projection effect.

\end{document}